# Spitzoid Lesions Diagnosis based on GA feature selection and Random Forest

Abir Belaala[1], Labib Sadek Terrissa[1], Noureddine Zerhouni[2], and Christine Devalland[3]

[1] LINFI Laboratory, University of
Biskra, Algeria
[2] FEMTO-ST Laboratory, Besançon, France
[3] Service of Anatomy and Pathology Cytology, Belfort, France

**Abstract.** Spitzoid lesions broadly categorized into Spitz Nevus (SN), Atypical Spitz Tumors (AST), and Spitz Melanomas (SM). The accurate diagnosis of these lesions is one of the most challenges for dermapathologists; this is due to the high similarities between them. Data mining techniques are successfully applied to situations like these where complexity exists. This study aims to develop an artificial intelligence model to support the diagnosis of Spitzoid lesions. A private spitzoid lesions dataset have been used to evaluate the system proposed in this study. The proposed system has three stages. In the first stage, SMOTE method applied to solve the imbalance data problem, in the second stage, in order to eliminate irrelevant features; genetic algorithm is used to select significant features. This later reduces the computational complexity and speed up the data mining process. In the third stage, Random forest classifier is employed to make a decision for two different categories of lesions (Spitz nevus or Atypical Spitz Tumors). The performance of our proposed scheme is evaluated using accuracy, sensitivity, specificity, G-mean, F- measure, ROC and AUC. Results obtained with our SMOTE-GA-RF model with GA-based 16 features show a great performance with accuracy 0.97, F-measure 0.98, AUC 0.98, and G-mean 0.97.Results obtained in this study have potential to open new opportunities in diagnosis of spitzoid lesions.

## 1 Introduction

Spitz nevus is a rare type of skin mole, it presents as a dome-shaped, pink to reddish-brown papule or nodule, typically less than 6 mm in diameter [1], usually affects young people and children. Spitz nevi are wholly benign; however spitzoid melanomas are malignant and have the possibility to spread to the regional lymph node basin and distant sites [2]. Atypical Spitz tumors (AST) are the most challenging, because of an uncertain malignant potential that has overlapping histologic features with Spitz nevus (SN) and spitzoid melanoma (SM). ASTs involve the sentinel lymph nodes at a greater frequency than conventional melanoma and often harbor chromosomal copy number changes, yet most cases follow an indolent course [3].

In recent years various computer-assisted diagnosis (CAD) systems on skin cancer have been proposed. Such systems provide a great assistance for diagnosing skin cancer lesions samples faster and more accurately. Most of research papers on AI in skin cancer are based on dermoscopic image analysis [4- 7]. However, the number of papers on spitz nevus is very limited due to two major reasons. One of the reasons is that this type of skin lesion is not common. The other reason is that historical data of diagnostics of spitz nevus is usually in the hands of doctors and it is usually hard for the scientific community to get. Our work uses advanced machine learning algorithms to classify this type of lesions not only based on dermascopic vision, but on clinical, histological, and immunohistochemistry features to make accurate diagnosis to distinguish between spitz nevus and atypical spitz tumors.

In this study, we develop a new scheme for spitzoid lesions diagnosis using clinical, histology, and immunohistochemistry dataset. The proposed scheme consists of the pre-processing phase, to improve data quality, amputate missing data, and solve data imbalance problem using over-sampling method SMOTE, in second phase , in order to improve classification performance, a feature selection mechanism based on genetic algorithm (GA) was used to select the important features. Finally, random forest classifier was applied to classify spitzoid lesions. The performance of our proposed scheme is evaluated using accuracy, sensitivity, specificity, G-mean, F- measure, ROC and AUC.

The organization of the rest of this paper is as follows. Section 2, describes our proposed method which consists of pre-processing phase, feature selection, and classification. Experimental results to demonstrate effectiveness of our system and a discussion are presented in Section 3 and finally the paper ends with conclusion in Section 4.

## 2   Proposed method

**Figure 1** shows the general schematic diagram of our proposed system. The details of each processing stage are described in the subsequent sections.

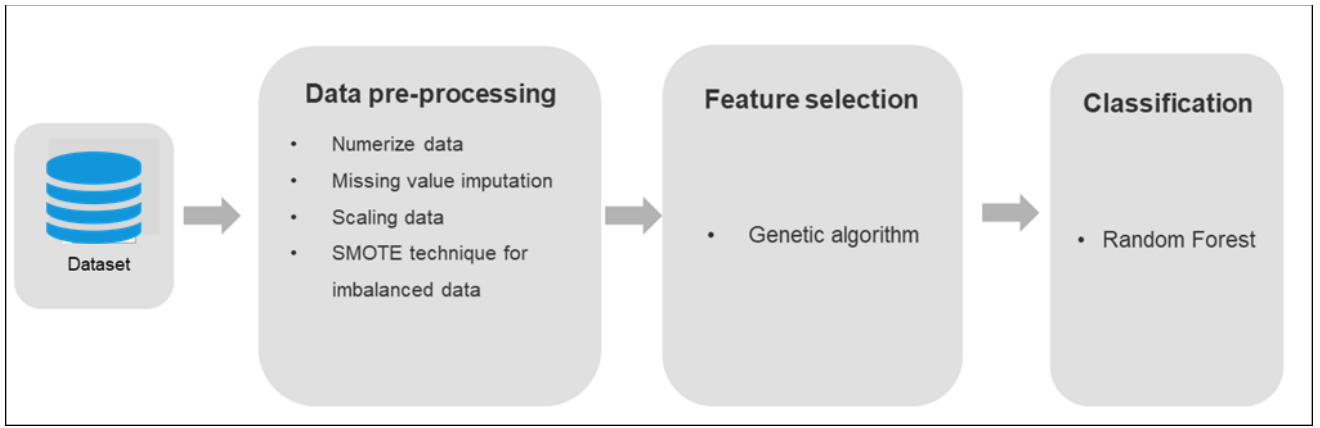

Figure 1: Genaral schema of our proposed process.

### 2.1   Data description

In this study, private data set from Pathology Department of Nord Franche-Comté Hospital (France) were used, it contains 54 patients, 45 have SN and 7 have AST. 29 attributes are calculated from clinical, histological, and immunohistochemical features. Details of the attributes are shown in Table 1.

| | | Feature | Input type | Input range | Details |
|---|---|---|---|---|---|
| Clinical data | 1 | Gender | Binary | 0 or 1 | Man : 0<br>Women: 1 |
| | 2 | Localization | quinary | 1 or 2 or 3 or 4 or 5 | 1:Trunk, 2:Lower extremity, 3:Upper extremities, 4:abdo 5:Face and neck |
| | 3 | Age | Continuous | From 2 to 54 | Majority of them are under 20 years old |

| | | | | |
|---|---|---|---|---|
| | 4 | Format | Ternary | 1 or 2 or 3 | 1: junctional, 2: wholly dermal. 3: compound |
| | 5 | Size of spitz | Continuous | From 0.3 to 1.4 | only 5 patients more than 1 cm, the rest under 1 cm |
| | 6 | Thickness | Continuous | From 0.1 to 6 | Majority < 2,5 mm |
| | 7 | Mitotic index | | From 0 to 2.2 | Majority < 0,5 per mm square |
| Histology data | 8 | Cytonuclear Atypia | Binary | 0 or 1 | 0: no<br>1: yes |
| | 9 | deep mitosis | | | |
| | 10 | Atypical Mitosis | | | |
| | 11 | Infiltration of the hypodermis | | | |
| | 12 | Asymmetry | | | |
| | 13 | Blurred boundaries | | | |
| | 14 | Pagetoid spread | | | |
| | 15 | Density of lymphocytic infiltrate | | | |
| | 16 | Hypercellularity | | | |
| | 17 | Ulceration | | | |
| | 18 | Kamino's body | | | |
| | 19 | desmoplastic cells | | | |
| | 20 | epidermal alteration | | | |
| | 21 | grenz zone infiltration | | | |
| | 22 | irregular nests | | | |
| | 23 | lack of maturation | | | |
| Immunohistochemistry | 24 | P16 | | | 100% no loss |
| | 25 | KI 67 | Continuous | From 0 to 18 | most of them < 5 |
| | 26 | BRAF | Binary | 0 or 1 | 0: mute, 1: not mute |
| | 27 | ALK IH | Binary | 0 or 1 | 0: negatif, 1: positif |
| | 28 | ALK Fish | Nul | Nul | Nul |
| | 29 | Melanin pigmentation | quaternary | 0 or 1 or 2 or 3 | |

Table 1: Spitz Nevus Dataset details.

## 2.2 Data Pre-Processing

In order to achieve more accurate results, data pre-processing is an important step for transforming raw Spitz nevus data into a clean and understandable format for analysis. In our dataset the majority of features are categorical (see Table 1). Since machine learning models are based on Mathematical equations we would only use numbers in the equations, for that we will convert them into numerical values. The second setup the missing values were replaced with the help of the estimated mean value. This method involves replacing a missing value with the overall sample mean [8]. It is easily implemented and simple. The third step is to solve the imbalance data, because our data is highly imbalanced, where we find 47 cases of classical Spitz nevus, and only 7 cases of Atypical Spitz Tumors. To solve this problem, SMOTE (Synthetic Minority Oversampling TEchnique) is implemented. This technique consists of synthesizing elements for the minority class, based on those that already exist [9]. It works randomly picking k a point from the minority class and computing the k-nearest neighbors for this point. The synthetic points are added between the chosen point and its neighbors. In our work we rescaled the columns to a range of [0-1], to avoid the numerical difficulties in calculation. [10].

## 2.3 Feature selection

The second phase in the Spitz nevus diagnosis process involves feature selection. In our case where there is too many attributes, this process is very important to find the attributes most relevant to the classification. In this work, we have used randomize wrapper method: Genetic algorithm (GA). GA have found wide range of applications [10,11,12], it operates with population, and the dominant solution is received after a sequence of iterative steps. GA develops sequential populations of periodic solutions that are presented by a chromosome until adequate results are reached [10]. Chromosomes are evaluated for their quality according to a predefined fitness function. Two major operators are crossover and mutation functions, and these have impact on the fitness value. Chromosomes for reproduction are selected by finding the fitness value, and the bigger fitness value is obtained, by selecting the chromosome with higher probability using either the roulette wheel or the tournament. In mutation, the genes may be changed randomly. The parameter settings for our Genetic Algorithm as feature selection are presented in Table 2, and Figure 2 shows the process of our GA based feature selection.

| Parameter | Value |
|---|---|
| Population size | 100 |
| Number of generation | 50 |
| Rate of crossover | 0.8 |
| Rate of mutation | 0.1 |
| Fitness evaluation | Accuracy of RF classifier |
| Size of chromosome | 27 |
| Coding | Binary<br>0: not selected<br>1: selected |

Table 2: parameter settings of our genetic algorithm for feature selection.

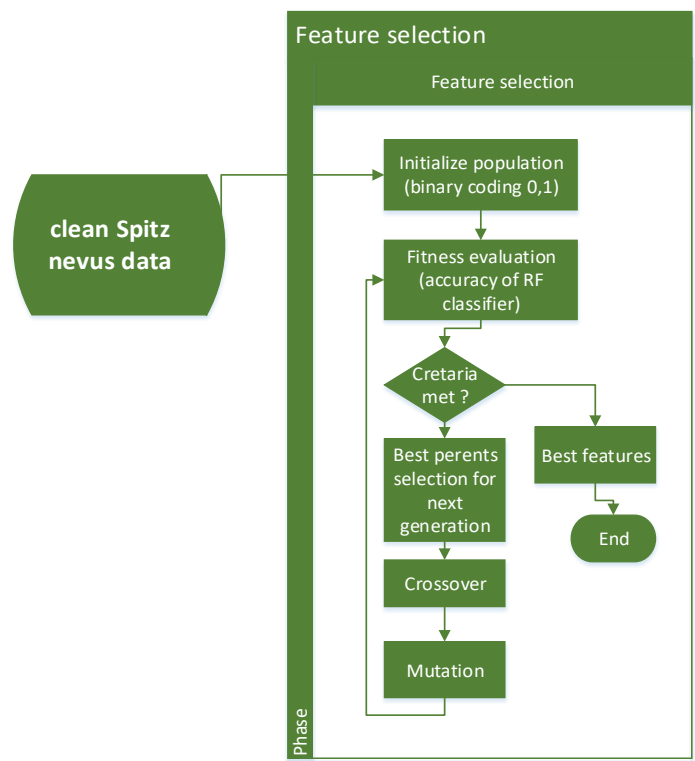

Figure 2: Process of our proposed GA based feature selection.

### 2.4 Classification

**Random Forest**

Random forest is defined by Breiman as "combination of tree predictors such that each tree depends on the values of a random vector sampled independently and with the same distribution for all trees in the forest" [13]. Random forest develops multiple decision trees as shown in Figure 3, to classify a new observation from an input vector, the observation is sent as input to each of the trees in the forest. Each tree specifies a classification, or "votes" for that class. At the end, the forest chooses the classification that has obtained the most votes among all trees [13]. A random forest is grown in three steps:

1. From the original training dataset S, RF select K samples randomly using bagging method

2. K samples are used as bootstrap sample for growing K decision trees using to achieve the K classification results

3. Each K classifier votes to elect the optimal classification with majority vote. If there are f input variables then a number k is specified such that at each node, k variables are selected at random out of the f features and the best split (using information gain) on these k is used to split the node.

RF does not overfit. It is fast and produces K number of trees for a large dataset in a few seconds. When the training set for the current tree is drawn by sampling with replacement, about one-third of the cases are left out of the sample and not used for bagging. This OOB (out-of-bag) data are used to get a running unbiased estimate of the classification error as trees are added to the forest. It is also used to get estimates of variable importance [14].

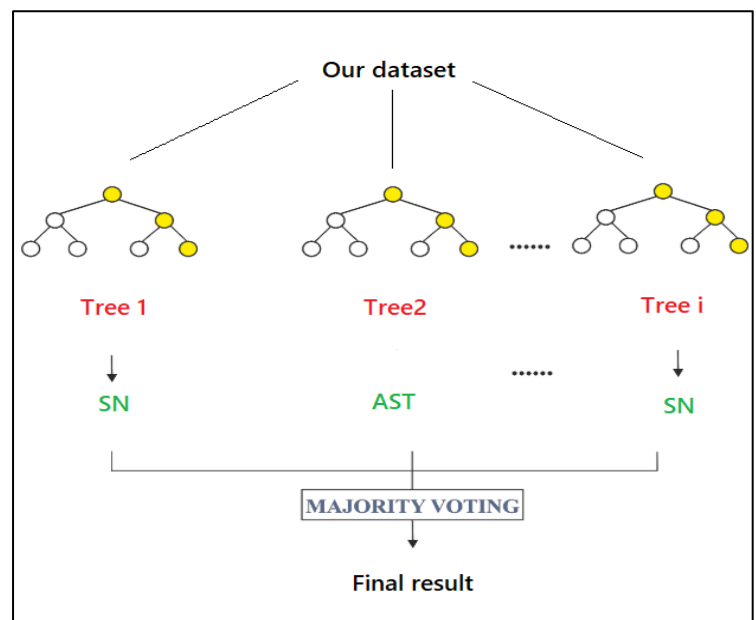

Figure 3: Working principle of RF

## 3 Results and discussion

This experiment is compiled and run in python language environment, and the default values of functions are used for all parameter values that are not explicitly stated. The 10-fold cross validation scheme was performed to evaluate the performance of the proposed method. **Table 3** shows the performance of Random Forest on our dataset without and with over-sampling method SMOTE, and GA based feature selection. In the results of RF without over-sampling method, RF's accuracy is high 0.85. Thus, we need to investigate RF's performance on other measures beyond accuracy. Among other measures, the sensitivity, specificity, and F-measure show a big difference between SN class which is very high, and minority class (AST) very low. We conclude that our data is imbalanced, where majority classes (SN) dominate over minority classes (AST), causing the RF classifier to be more biased towards majority classes. This causes poor classification of minority classes. For this reason, we applied SMOTE over-sampling method to solve this problem. The input parameter that need to be determined in SMOTE method is number of nearest neighbors k. we tried several different k values in the experiment, finally decide to use k = 6 which gives best accuracy. We can observe that performance of RF has been improved after using SMOTE, where we notice a balance in sensitivity, specificity, and F-measure of both SN and AST class. Moreover, it gives a higher accuracy 0.95 and G-mean 0.95. These results suggest that SMOTE method is a good technique to balance our Spitz nevus dataset, Figure 4 shows the same results using ROC curve, we notice a great improvement when we applied SMOTE technique from AUC=0.55 to AUC = 0.92.

In the second test, we first used genetic algorithm (GA) based feature selection to select the best attributes, and FR accuracy as fitness evaluation. Then, we used the same machine learning classifier for classification. Experimental results showed that highest classification performances are achieved when GA is used as feature selection. It gave highest accuracy 0.97, F-measure 0.98, and G-mean 0.97 with only 16 selected features (see **Table 4**), and AUC = 0.98 as shown in **Figure 4**.

|  | Accuracy | Sensitivity | | Specificity | | F1-measure | | G-mean |
|---|---|---|---|---|---|---|---|---|
|  |  | AST | SN | AST | SN | AST | SN |  |
| FR Without Over-sampling methods | 0.85 | 0.14 | 0.96 | 0.33 | 0.88 | 0.20 | 0.92 | 0.85 |
| RF with SMOTE k=6 | 0.95 | 0.94 | 0.94 | 0.98 | 0.98 | 0.96 | 0.96 | 0.95 |
| RF-SMOTE with GA | 0.97 | 1.00 | 0.96 | 0.96 | 1.00 | 0.98 | 0.98 | 0.97 |

Table 3: Experimental performance on our proposed method.

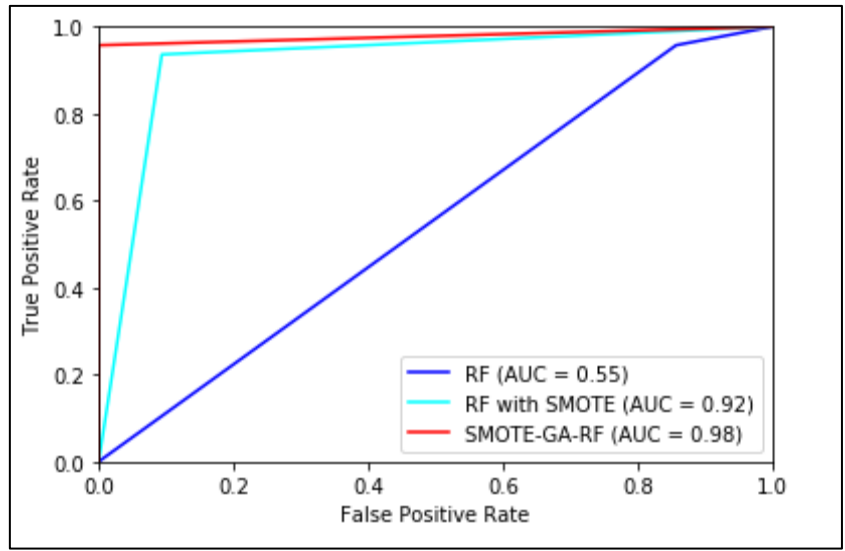

Figure 4: ROC curve of our proposed method.

| GA-RF | | | |
|---|---|---|---|
| Gender | yes | Kamino's | no |
|  |  | body |  |
| Localization | yes | Desmo | no |
| Age | yes | Modif epid | yes |
| Format | no | Grenz | no |
| Diameter | yes | Th irreg | no |
| diameter >1 | no | No grad | yes |
| Thickness | no | KI 67 | yes |
| Mitotic index | yes | Ki67>1% | yes |
| Cytonuclear Atypia | yes | BRAF | no |
| mitoses profaned | no | ALK IH | yes |
| Infiltration of the hypodermis | yes | Ly | no |
| Asymmetry | yes | Pig mél | yes |
| Blurred boundaries | no | Pagetoide migration | yes |
| hyperCELL | yes |  |  |

Table 4: Selected features by SMOTE-GA-RF.

## 4 Conclusion

The suggested system accomplished higher classification accuracy rate, by improving the data quality, solving class-imbalance problem, decreasing the number of attributes and obtained higher performance rate, and identify the most important features that can give dermatologistes an idea of the features to be used in distinguishing spitzoid lesions. The SMOTE-GA-RF model obtained from this artificial intelligence method can be used as a medical decision support system for dermapathologists to make accurate classification with lower time, cost, and effort. In our future work, we aim to collect more data and add a diagnosis of Spitz Melanomas (SM), which is a third class of Spitz lesions. In addition to this, diffrent hybrid classifiers can be applied to our dataset in ordre to improve the performance of the proposed model.